\pdfoutput=1
\documentclass[acmsmall,nonacm,screen]{acmart}

\usepackage{booktabs}
\usepackage{amsmath}
\usepackage{tikz}
\usetikzlibrary{shapes, arrows, positioning}
\usepackage{verbatimbox}
\usepackage{listings}
\usepackage{graphicx}
\usepackage{spverbatim}
\usepackage{tabularx}

\newcolumntype{L}{>{\raggedright\arraybackslash}X}%

\usepackage[linesnumbered,lined,commentsnumbered]{algorithm2e}

\SetCommentSty{mycommfont}

\usepackage{caption}
\begin{document}

\title{A Generative Neural Network Framework for Automated Software Testing}
\author{Leonid Joffe}
\affiliation{
	\institution{UCL}
	\country{UK}
}
\email{leonid.joffe.14@ucl.ac.uk}

\author{David J. Clark}

\affiliation{
	\institution{UCL}
	\country{UK}
}
\email{david.clark@ucl.ac.uk}



\begin{abstract}

Search Based Software Testing (SBST) is a popular automated testing technique which uses a feedback mechanism to search for faults in software.
Despite its popularity, it has fundamental challenges related to the design, construction and interpretation of the feedback.
Neural Networks (NN) have been hugely popular in recent years for a wide range of tasks.
We believe that they can address many of the issues inherent to common SBST approaches.
Unfortunately, NNs require large and representative training datasets.

In this work we present an SBST framework based on a deconvolutional generative neural network.
Not only does it retain the beneficial qualities that make NNs appropriate for SBST tasks, it also produces its own training data which circumvents the problem of acquiring a training dataset that limits the use of NNs.

We demonstrate through a series of experiments that this architecture is possible and practical.
It generates diverse, sensible program inputs, while exploring the space of program behaviours.
It also creates a meaningful ordering over program behaviours and is able to find crashing executions.
This is all done without any prior knowledge of the program.

We believe this proof of concept opens new directions for future work at the intersection of SBST and neural networks.

\end{abstract}


\maketitle

\section{Introduction}

In this paper we explore an automated testing framework based on Neural Networks (NN).
The proposed approach aims to address some of the most pertinent problems of both automated testing and generative NN models.
This proof of concept introduces multiple new ideas at the intersection of automated software testing and generative neural networks, and we hope it will stimulate further research in this area.

Automated testing techniques have become increasingly popular thanks to the availability of resources and their ever improving effectiveness.
We view automated testing from the perspective of Search Based Software Testing (SBST) \cite{harman2001search, mcminn2004search}.
In SBST, a program is repeatedly executed, its execution monitored and further executions are generated with the aim of more effective fault discovery.
Commonly the goal is to optimise coverage.
A fundamental feature of SBST is the reliance on a feedback mechanism to evaluate and direct the search.

The practical instantiation of SBST that we consider is fuzzing.
Fuzzing is a technique where a program is bombarded with random inputs in hopes it will eventually crash \cite{sutton2007fuzzing}.
Modern fuzzers however use feedback mechanisms to improve their effectiveness.
Using a feedback loop for improving search brings fuzzing into the realm of SBST.
A popular modern fuzzer is the American Fuzzy Lop (AFL) \cite{zalewski2007american}.
Although it does use a feedback mechanism and hence falls into the realm of SBST, its search strategies and notions of similarity are non-principled heuristics -- they "just work".
That said, these heuristics all ultimately drive the fuzzer towards exploring a program's behaviours maximally \textit{diversely}.
Indeed, diversification is generally a very common target for testing \cite{ammann2016introduction, heimdahl2004specification, gay2015risks};
after all, if the target of search is unknown, the best you can do is explore.

Although popular both in academia and industry, SBST is not without its limitations \cite{mcminn2011search, aleti2017analysing}.
The problems of SBST we aim to address are the following.
\textit{First}, fitness landscapes of SBST may have plateaus or be discontinuous.
Then either multiple adjacent candidate solutions appear equivalent in terms of fitness and the search mechanism cannot prioritise them or it cannot move to a better solution.
\textit{Second}, the fitness landscape may contain local optima which leads the search to a sub-optimal solution.
\textit{Third}, choosing the representation requires domain knowledge and expert involvement \cite{shepperd1995fundamentals}.
\textit{Fourth}, it is not apparent how to assign an ordering onto the search landscape, likewise requiring an involvement of an expert \cite{shepperd1995fundamentals, harman2004metrics}.
Granted, it can be defined in terms of the search operators, i.e. candidate solutions one search step away from each other are adjacent.
But is this the best ordering for the search?
\textit{Finally}, the generation of new candidate solutions is a big can of worms with various approaches and solutions \cite{mcminn2004search, anand2013orchestrated, ali2010systematic, alshraideh2006search, fraser2013whole, fraser2011evosuite, fraser2012mutation, pacheco2005eclat, korel1992dynamic}.
What search operators to use?
How much prior knowledge is required and available?
How to produce the next candidate solution?

We propose that NNs' properties make them ideal for tackling these issues.
\textit{First}, NNs are trained by a process of backpropagation \cite{rumelhart1986learning} which means they must be differentiable and thus continuous by construction \cite{glasmachers2017limits}.
That is, if a neural network trains, its intermediate states must be differentiable.
This continuity and differentiability make NNs a natural candidate to tackle the issue of plateaus in SBST search spaces.
\textit{Second}, it has been shown that given sufficient size, NNs avoid local optima \cite{kawaguchi2016deep, swirszcz2016local, nguyen2017loss, nguyen2018optimization}.
\textit{Third}, the above property also implies that if a representation contains a useful signal, an NN will discover it.
This means that they can use representations that are difficult to interpret manually.
NNs may suffer from noise given redundant data, but these problems can be addressed with feature selection \cite{verikas2002feature,leray1999feature, wang2014attentional} and modern, deep architectures suffer less from this problem \cite{li2018feature}.
This alternative is nonetheless preferable to a major manual effort.
\textit{Fourth}, due to their differentiable nature, NNs impose an order relation onto data (e.g. \cite{kingma2013auto}).
They may thus help us reason about similarity and diversity of program behaviour in a principled, continuous way.
\textit{Lastly}, NNs can be used for generating new data without analytical human effort \cite{Goodfellow-et-al-2016, kawthekarevaluating, chollet2017git, graves2013generating, openai_gnn, radford2015unsupervised, van2016wavenet}.

NNs, as tools for SBST, come with limitations of their own however.
The main problem is that they are data hungry; they need large representative training datasets \cite{beleites2013sample}.
This appears like a disqualifying issue in the context of SBST.
If one wishes to train an NN to be used as a fuzzer, and has sufficient data to train the NN, this data could simply be used to test the program itself.
This really defeats the purpose of building and training an NN-based fuzzer.

The architecture proposed in this work can address all these problems -- SBST and NN related alike.
It is a generative model that produces its own training data.
In a way, this architecture is similar to that of reinforcement learning (RL) where an agent explores an environment, discovers rewards and learns to navigate the space more effectively.
In our approach, the agent is the generative model while the program under test is the environment.
This might seem like a bizarre proposition as there is no apparent way to evaluate the quality of the generated data.
You can produce all the random data you like, but how do you know if it is any good?
What is your reward signal?
We suggest that the principle of diversification can be adapted from SBST to evaluate the generated data;
the model is rewarded for \textit{diversity}.

This framework we call GNAST (Generative Network for Automated Software Testing) begins by throwing random inputs at a program.
Most of these will be rejected by the program.
Some, however, will trigger an unusual execution trace.
Those are prioritised and kept in the training dataset.
As the process continues, GNAST generates program inputs that trigger new behaviours and uses them in its training dataset.
The fact that GNAST is NN based allows it to address the issues outlined above, as we will show in the sequel.

In this work we implement a prototype of GNAST and present a number of initial findings.
First and foremost, we show how such a system can be trained and how it produces diverse program inputs.
Furthermore, the inputs are clearly sensible with respect to the syntax of the program under test.
In addition, we can control the syntactic similarity of generated strings.
Finally, rudimentary as the current state of GNAST is, it does actually discover crashes.
Currently it is a prototypical proof of concept accompanied by several outstanding questions.
As such, it cannot be readily compared with fully fledged fuzzers like AFL.

Even though it is a prototype, GNAST presents a number of novel ideas.
It is a generative NN-based automated software testing tool that does not require a training dataset -- it produces its own.
It is a new example of a deconvolutional generative model for string generation.
It uses a novel quantified notion of similarity for program executions.
It presents a prioritisation method for program executions based on a greedy algorithm called Farthest-First Traversal (FFT).
Finally, it uses "unusualness" and diversity as an explicit training target and although this idea is common in testing, it is novel in this formulation.

\section{Overview of Approach}\label{sec:overview}

We propose a framework for automated software testing.
Its purpose is to explore the behaviours of a Software Under Test (SUT) diversely.
Our system does this by generating inputs for the SUT, observing the executions under those inputs and adjusting further generation of inputs towards the most unusual behaviours.
It is essentially an evolutionary fuzzer, albeit with convoluted\footnote{Pun intended.} generative and feedback mechanisms.

\begin{figure}
	\centering
	\includegraphics[width=\linewidth]{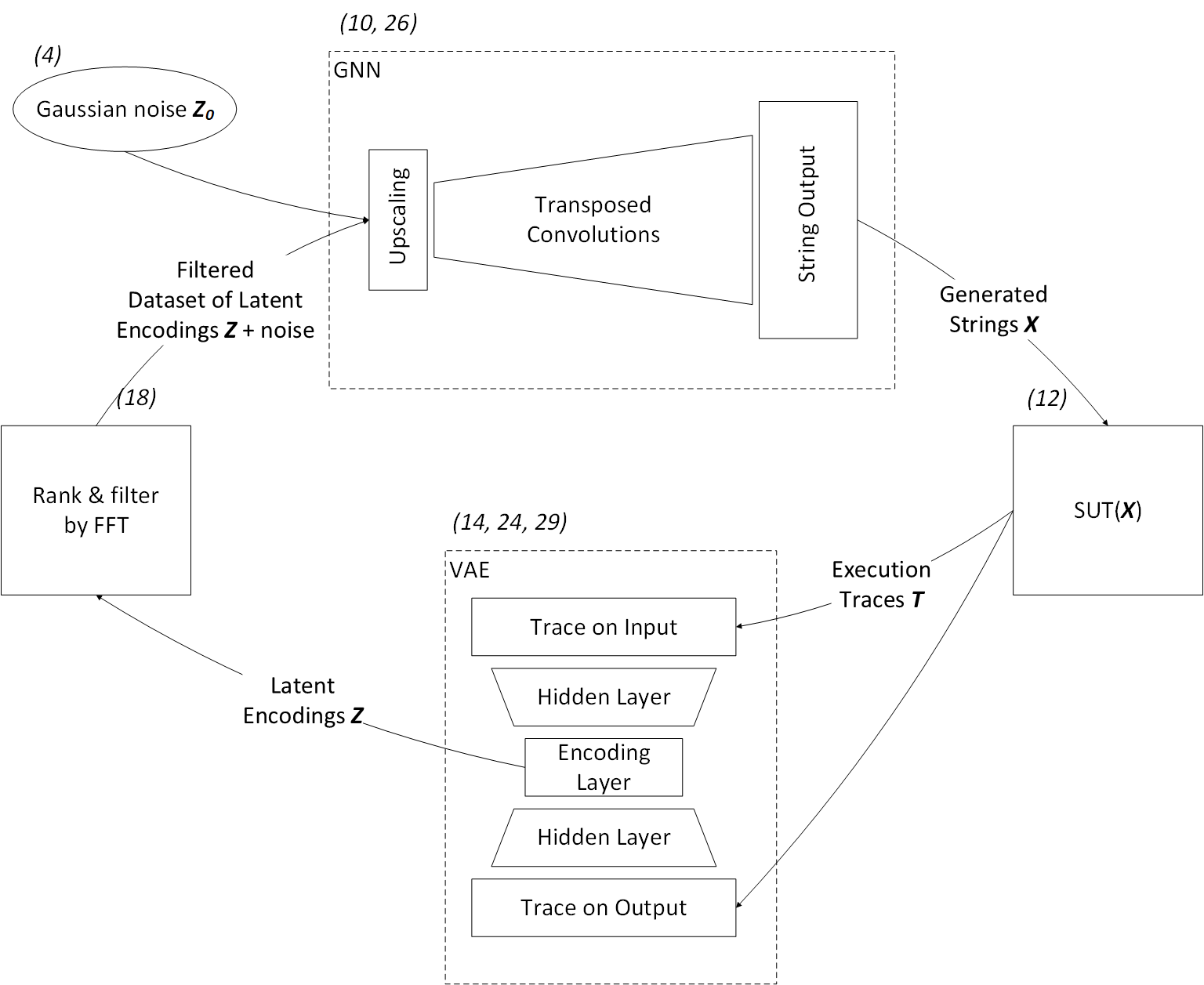}
	\caption[GNAST Framework]{The GNAST Framework. A detailed explanation of the training and generative processes is given in \autoref{sec:overview}. The algorithm corresponding to this image is shown in \autoref{alg:gnast}, with numbers in brackets corresponding to line numbers.}
	\label{fig:framework}
\end{figure}

The tool is based on two neural networks (NN), an execution trace profiler and a prioritisation mechanism of executions.
The structure is shown in \autoref{fig:framework} and its algorithm is presented in \autoref{alg:gnast}.
A single epoch of the algorithm corresponds to two passes through the framework in \autoref{fig:framework}:
the first is a generative pass where the networks' weights are not updated, and a training pass -- where they are.
The algorithm and framework are described next, with the numbers in brackets corresponding to the lines in \autoref{fig:framework} and \autoref{alg:gnast}.

The process is initialised by feeding Gaussian noise to an untrained Generative Neural Network (GNN) (\textit{2-6}).
It produces a batch of program inputs $X$.
Most of these will be nonsensical; strings of random characters.
The inputs are then executed by the SUT and the execution traces $T$ are collected (\textit{12}).
As most inputs are just noise, most execution traces will yield the "invalid input" execution trace.
Some inputs however may contain features that are valid, which will be reflected in their execution traces.

The execution traces are then encoded by a Variational Autoencoder (VAE) (\textit{14}).
It casts the discrete, unorderable execution trace into an n-dimensional "latent" space encoding $E$.
The latent space encoding is a quantifiable representation of features of execution traces, and we can reason about their similarity in terms of Euclidean distance.
The encoded executions are then ranked by the "unusualness" of their encoded traces using the Farthest-First Traversal (FFT) algorithm (\textit{18}).
Redundant datapoints are discarded and the most interesting ones are kept in a persistent training dataset (\textit{20}).
The dataset is composed of the execution traces $T$, their encodings $E$ and the program inputs that triggered them $X$.

The dataset is then used to train both the VAE and the GNN (\textit{24, 26}).
As the VAE learns to encode the execution traces of the training dataset, new, unusual ones stand out from the bunch.
The GNN, in turn, learns to produce program inputs that trigger a variety of traces, i.e. program behaviours.
We also add noise to perturb the dataset towards exploration so that more novel behaviours are found (\textit{29}).
We suggest that the proposed framework represents a fundamentally novel approach to using neural networks for diversity driven testing of programs.

\begin{algorithm}
	\tcp{Initialise epoch counter}
	$e \leftarrow 0$\;
	\tcp{Initialise inputs to GNN from Gaussian noise}
	$\{Z_0\} \sim \mathcal{N}(\mu,\,\sigma^{2})$\;
	
	\tcp{Initialise dataset of program inputs, traces and GNN inputs}
	$\{Res_0\} \leftarrow \{<X=\emptyset, T=\emptyset, Z=Z_0\}$ \;
		
	\While{$\top$}{

		\Begin(Generative pass)
		{
			\tcp{Generate a batch of program inputs}
			$\{X_e\} \leftarrow GNN(Z_e)$\;
			\tcp{Execute SUT, collect traces}
			$\{T_e\} \leftarrow SUT(X_e)$\;
			\tcp{Get encodings of traces}
			$\{Z_e\} \leftarrow VAE_{train=\bot}(T_e)$\;
			
			\tcp{Append new inputs, traces and strings to the dataset}
			$\{Res_{e-1}\} \cup \{<X_e,T_e,Z_e>\}$
			
			\tcp{Rank traces with FFT}
			$R_e \leftarrow FFT(T_e)$\;
			\tcp{Discard redundant datapoints, limit of k=5000 due to hardware resource limitation}
			$\{Res|r_i<k\}$\;			
		}
	
		\Begin(Training pass)
		{
			\tcp{Train VAE on representative traces}
			$VAE(T_e)_{train=\top}$\;
			\tcp{Train GNN on representative noises and inputs}
			$GNN(<X_e,Z_e>)_{train=\top}$\;
		}
		\tcp{Update the inputs with encodings of traces}
		$\{Z_e\} \leftarrow VAE(T_e)_{train=\bot} + \epsilon \sim \mathcal{N}(\mu,\,\sigma^{2})$\;
		\tcp{Increment epoch}
		$e \leftarrow e+1$\;
	}
	
	\caption{The algorithm for training and generating inputs with GNAST. The algorithm corresponds to the image in \autoref{fig:framework} and described in \autoref{sec:overview}.}
	\label{alg:gnast}
\end{algorithm}

\section{Research Questions}

While many of the individual mechanisms of GNAST are inspired by other work, the overall structure is fundamentally novel.
It is an RL-inspired loop that generates training data for itself by sampling the output distribution and evaluates samples with an external diversity-driven oracle.
The novelty brings about a huge number of design and configuration decisions, all of which have an effect on the research questions outlined below. \\

\textbf{First}, there was no guarantee that the training of such a system would converge at all.
It is also not clear what optimisers, layer sizes, numbers of hidden layers etc. to use.
Non-convergence is essentially underfitting -- the mechanism does not learn to approximate the data.
Whether GNAST's training converges is the the focus of the first research question. \\
\textbf{RQ1}: \textit{"Does GNAST framework's training converge?"} \\

\textbf{Second}, it is insufficient for GNAST's training to simply converge.
It also needs to \textit{not} converge too far, so as to continue generating new datapoints.
New datapoints are essential both for exercising varied behaviours of the SUT as well as building up a diverse dataset for training.
Much like non-convergence in RQ1 means underfitting, converging too far corresponds to overfitting -- the system learns to produce only a few datapoints and since those are kept in the training dataset, their effect becomes ever stronger.
The second research question looks at diversity during training. \\
\textbf{RQ2}: \textit{"Does GNAST maintain diversity throughout training?"} \\

\textbf{Third}, if the mechanism does train in an acceptable way, we then need to evaluate whether the produced program inputs are sensible.
Granted, a program may crash under a completely unexpected random input, but a fundamental principle of fuzzing (and indeed any other testing) is that inputs ought to be consumable by the SUT, beyond an "invalid input" check.
This means they ought to be somewhat well-formed, or at least have some relevant syntactic features.
AFL generates strings that can hardly be called well-formed, as the representation yielded by its instrumentation is only a crude representation of a program's behaviour.
Since the instrumentation is lifted out of AFL, the strings generated by GNAST were expected to contain some syntactic feature similar to those made by AFL, but there was \textit{no} expectation of the them being properly well-formed.
The aim of the third research question is to see whether the generated strings are completely random or comparable in structure to those made by AFL. \\
\textbf{RQ3}: \textit{"Do the generated program inputs have syntactic features similar to those generated by an AFL baseline?"} \\

\textbf{Fourth}, one of the intended features of GNAST is the ability to control the similarity of syntactic features of the produced inputs by adjusting the input.
Once GNAST is trained, the latent space that was used as input to the GNN can be replaced with n-dimensional normal noise.
The GNN thus becomes a stand-alone generator which takes a vector of reals as input and generates a string on the output.
Input values close to each other are ought to produce similar strings while distant inputs should produce dissimilar ones.
Whether this is the case is investigated in the fourth research question. \\
\textbf{RQ4}: \textit{"Do strings generated from nearby points in the latent space share syntactic features vs. those far apart?"} \\


\textbf{Fifth}, by the same design as syntactic similarity of generated program inputs, GNAST ought to be able to generate strings that would trigger a specific desired behaviour.
After all, one of the components of the loss function is the reconstruction of input.
Our fifth research question looks at this aspect. \\
\textbf{RQ5}: \textit{"Can GNAST generate input strings that trigger specific behaviours of a SUT?"} \\

\textbf{Finally}, although GNAST is only a proof of concept with a lot of additional work to be done, we would like to know if the framework "has legs" as a design for a fuzzer.
The ultimate aim of a fuzzer is to find crashes, and our last research question is simple but poignant. \\
\textbf{RQ6}: \textit{"Does GNAST discover crashes in \textbf{sparse}\footnote{The program in which AFL found crashes."}?}


\section{Related Work and Background}

The proposed framework applies machine learning techniques, specifically GNNs, to automated software testing.
This section links the mechanisms of GNAST to concepts, ideas and inspirations in those fields.

\subsection{Fuzzing and the American Fuzzy Lop}

Fuzzing is an automated testing technique which attempts to discover faults by bombarding a program random inputs \cite{sutton2007fuzzing}.
Modern fuzzers use feedback to improve the input generation process;
they evaluate their progress in order to search for faults more effectively.

American Fuzzy Lop (AFL) is a popular fuzzer widely used in academia \cite{zalewski2007american}.
A number of recent papers have taken AFL as a basis, and improved on it.
These include work on improving AFL's instrumentation \cite{gan2018collafl, chen2018angora}, alternative search strategies \cite{bohme2017coverage, bohme2017directed, petsios2017slowfuzz}, producing better initial seeds \cite{lv2018smartseed}, locating interesting input string mutation locations \cite{rajpal2017not, she2018neuzz} and introducing program context information \cite{rawat2017vuzzer}.
Although we use some parts of AFL, our work does not attempt to improve AFL.
Instead, we are proposing a fundamentally different architecture.

\subsection{Representation and Fitness Function}
Feedback-driven automatic testing tools like modern fuzzers, fall under the field of Search Based Software Testing (SBST) \cite{harman2012search, mcminn2004search}.
As any other SBST approach, the design of a fuzzer depends on a representation and a fitness function \cite{harman2001search}.

Representation is the choice of what to observe about a candidate solution, in this case an execution.
It may be a coverage profile, a sequence of executed basic blocks or something simple like execution time.
In this work, the representation is based on the execution trace profiling mechanism from AFL.
AFL's representation of an execution is an approximate count of decision point transitions.
It does not capture context nor sequence information about the execution.

The fitness function is what transforms a representation into a quantified assessment of quality of a candidate solution.
But what constitutes quality in fuzzing and how can it be quantified?

Fuzzing, and software testing in general, aims to exercise program behaviours diversely:
diversity is quality.
The original AFL uses a handful of heuristics to identify executions that are distinct or "interesting", i.e. of high quality.
For instance, if an execution exercises a transition that has not been previously seen, it is considered interesting.
Whilst the heuristics of AFL are empirically effective, they are not principled and do not order nor quantify the similarity of executions or behaviours.

\subsection{Quantifying Program Behaviour}

GNAST is intended to give an ordering, and quantify the similarity of executions.
The approach relies on an autoencoder neural network which is used to process execution traces.
An autoencoder is an NN which attempts to reconstruct its inputs at the outputs, while arranging the datapoints into an n-dimensional encoding ("latent") space by similarity of their features \cite{Goodfellow-et-al-2016}.
Specifically, we use a flavour of an autoencoder called the Variational Autoencoder (VAE) \cite{kingma2013auto}.
The essential detail of a VAE is in the structure of its encoding layer.
Rather than encoding datapoints as single real number values, they are instead encoded as tuples of mean and variance $(\mu, \sigma^2)$.
The result of this construction is that datapoints tend to be more evenly spread across the latent space and interpolation between them is smoother than in a regular autoencoder.

In GNAST, the inputs and outputs to the VAE are the execution traces and the latent space encoding is a representation of their most salient (i.e. most distinguishing) features.
The benefit of encoding traces in this way is that it imposes an order relation on discrete, seemingly unorderable datapoints.
Furthermore, the space is continuous and smooth.
These are characteristics of a good search landscape \cite{harman2004metrics}.
Since execution traces are the representation of program behaviour, the latent space in fact constitutes the space of behaviours.
We suggest that since we want to explore diverse behaviours, this latent space is what we ought to be exploring diversely.


\subsection{Ranking Algorithm}

Once we have constructed an n-dimensional Euclidean space to represent program behaviours, we need a method for evaluating which datapoint is unusual and which is redundant.
We propose doing this using an algorithm known as the Farthest-First Traversal (FFT) or the greedy permutation.
FFT arranges a set of points into a sequence such that the minimal distance of each following point is maximised.
In other words, the prefix of the sequence is always maximally representative of the whole dataset.
We are not aware of work which would use FFT in ranking candidate solutions in SBST.

The algorithm is shown in \autoref{alg:fft} and described below, with line numbers shown in brackets.
Prior to the actual algorithm, the pairwise distance of each datapoint is calculated (\textbf{2}).
Then the result is initialised with two maximally distant elements (\textbf{3}).
The next element to append is chosen such that it is furthest from ones already in the result sequence, i.e. the minimal distance is maximised (\textbf{5}).
Each following element in the sequence is thus maximally different from ones already in the prefix.

Once the dataset is sequenced, its tail (the most redundant portion) is discarded.
We propose that pruning the dataset according to this notion of similarity keeps it diverse and maximally representative.


\begin{algorithm}
	\KwData{List \textit{X} of n-dimensional coordinates of elements \textit{x}}
	\KwResult{Result \textit{Y} of tuples \textit{y} of (indices, distances) $(i,d)$}
	\Begin{
		Calculate the pairwise distance matrix $D$\;
		Initialise \textit{Y} with $y_0, y_1$ of $d_{max}$\;
		\While{$|Y|<|X|$}
		{
			Append $x$ s.t. \textbf{min} distance to $\forall y \in Y$ is \textbf{max}\;
		}
		\KwRet{$(i,d)\in Y$}
	}
	
	\caption{The greedy permutation or Farthest-First Traversal (FFT) sequences elements such that each successive element maximises the minimal distance to elements in the result.}
	\label{alg:fft}
\end{algorithm}

%




\subsection{Generative Neural Networks}\label{sec:gnn}


The heart of GNAST is a Generative Neural Network (GNN) that is trained to produce program inputs using the culled, representative dataset.
As the dataset expands, the generator gets new datapoints to drive its training towards previously unseen behaviours and program inputs.

Perhaps the best known example of a GNN design for strings is the Recurrent Neural Network (RNN) \cite{Goodfellow-et-al-2016, rumelhart1986learning, hochreiter1997long}.
A very early version of GNAST was implemented with an RNN (modelling on work by Bowman \textit{et al.} \cite{bowman2015generating}), but it was much too slow to train and to generate new inputs, so that design was abandoned.

Our generator is therefore based on an alternative design, inspired by the Deconvolutional Generative Adversarial Network (DCGAN) \cite{radford2015unsupervised}.
In DCGAN, two neural networks -- a generator and a discriminator -- are pitched against each other.
The generator takes Gaussian noise as input, passes it through a stack of transposed convolutions \cite{Goodfellow-et-al-2016} and produces an image on the output.
The discriminator takes an image as input and outputs an answer of whether the image was real or fake (generated).
While the discriminator learns to be ever more effective in telling fake images from real ones, the generator attempts to produce images that fool the discriminator into believing they are real.
This tandem of networks ends up creating realistic images.

There are two significant problems of readily applying DCGAN to generation of discrete data.
The first problem is that unlike pixel values in images, discrete variables such as characters in strings are not readily orderable.
For instance, "bat" is \textit{not} semantically almost the same as "cat" although the two words only differ by a single (alphabetically adjacent) character:
$"cat" \neq ("bat"+1.0)$ \cite{goodfellow_reddit_2016}.
On the other hand, in continuous variables such as pixel hue values, a slight offset does not dramatically change the meaning:
a cat can be still identified even if the image is slightly blurry.

The other problem, inherent to neural networks, is the need for a lot of training data \cite{beleites2013sample}.
DCGAN requires a corpus of real datapoints for training the discriminator.
This in itself is a disqualifying requirement for a fuzzer, as it would require a large, representative dataset of valid program inputs.
This defeats the purpose of a fuzzer:
if you have a truly representative dataset of inputs, it could just be used to test the program.
Furthermore, using only valid inputs would train the generator to produce only valid inputs.
This is not ideal for fuzzing, as the program ought to be tested on both valid and invalid inputs.

In GNAST, in place of a discriminator network, we have the combination of a SUT, a VAE and FFT ranking.
This mechanism serves the same purpose as the discriminator in DCGAN: to, as it were, quantify the quality of candidate solutions.
Furthermore, thanks to the VAE, we have a continuous input space ordered by features of execution traces.
This ordering of the input space imposes an order on the generated strings.
In addition, we have unlimited training data thanks to GNAST producing its own.
These design features thus address major problems of the use of generative models for string generation.

In the machine learning arena, the networks employed in this prototype are not novel.
Recent advances in GNNs have brought about numerous novel architectures, many of them for image generation, e.g. \cite{he2016identity, karras2019style, zhang2017stackgan}, and even for text generation \cite{bowman2015generating, wang2018text}.
Indeed the concept of training an NN given some response from an external source is reminiscent of RL.
From this point of view, the GNN represents the actor, the SUT is the environment, and the VAE with FFT is the critic\cite{konda2000actor}.

As for the direct use of GNNs for test input generation -- the closest use case to ours -- we are only aware of work by Godefroid et al. where an RNN was used to test a PDF reader \cite{godefroid2017learn} with mixed results.

\section{Experimental Setup}\label{sec:experiments}

This section presents the implementation details of each component of GNAST, as well as the experiments we ran.
To the best of our knowledge, the proposed framework is novel in many aspects.
A large part of our experiments was therefore exploratory;
various designs and configurations were considered and tested, albeit not exhaustively due to resource limitations.
The NN mechanisms were implemented with the Tensorflow framework \cite{tensorflow2015-whitepaper}.
Citations to standard deep learning concepts are omitted for clarity, and the reader is referred to the Deep Learning Book \cite{Goodfellow-et-al-2016} for reference.

\subsection{Trace Profiling}

Two mechanisms are lifted out of AFL:
the program instrumentation (and hence representation) and the execution harness.
To be clear, \textit{none} of AFL's generative or prioritisation mechanisms were used.
We did not modify or parametrise these mechanisms, so there are no options to be considered, save for, of course, replacing the instrumentation with an alternative tool altogether.

AFL's execution trace representation is a count of decision point transitions.
First, the source code is instrumented with a drop-in replacement for GCC (or G++ or Clang) prior to testing.
For efficiency, only a rough number of transitions is kept (8 buckets), and the trace is of a fixed size of 64K.
An execution trace is thus a 64K long, typically very sparse vector with 8 distinct values.
Although GNAST does not take advantage of the efficiency of this representation but it is nonetheless appropriate for three reasons.
First, it is simple to implement by adapting AFL's code.
Second, the empirical evidence of AFL's performance is in and of itself an indication of the representation's effectiveness.
Finally, there are many new aspects to our framework, and using an alternative, potentially ineffective trace representation would add to the uncertainty of the framework overall.

AFL's execution harness is referred to as a "fork server".
The basic idea of the fork server is to spin up the SUT up to the initial \verb|main()| call and keep a snapshot of the program in that state.
Whenever the SUT is executed under a new input, this initial state is copied and execution proceeds from there.
Thanks to this simple mechanism, the time required for initialisation is avoided and fuzzing becomes much faster, so the inputs produced by the fuzzer can be executed quickly.

\subsection{Variational Autoencoder}

The VAE is composed of an encoder and a decoder, modelled on the architecture of Kingma's work \cite{kingma2013auto}.
The encoder is as follows.
The first layer is the input of size 64K, matching the size of the trace.
This is followed by one or more densely connected hidden layers.
Next is the encoding layer which is composed of a densely connected $\mu$ and $\sigma^2$ layers.
In addition, the encoding layer includes a source of Gaussian noise.
The construction of the encoding layer $z$ is shown in \autoref{eq:encoding_layer}.
Once the VAE has been trained, the VAE is queried for an encoding of a trace, it is $\mu$ layer's output that is fetched.
That is, $\sigma^2$ and $z$ layers are only used during training to give the latent space the desired characteristics, but the actual encoding is the value $\mu$.

\begin{equation}
	z = \mu + e^{0.5\sigma^2} * \mathcal{N}(0,I)
	\label{eq:encoding_layer}
\end{equation}

The output of the $z$ layer is fed to hidden layer(s) of the decoder.
The output of the decoder a categorical softmax layer of size $(64K, 8)$.
This is specific to this trace representation where counts of state transitions are bucketed into 8 values.
This allows the VAE to be trained using categorical cross-entropy.

The loss function of the VAE is composed of two terms, as shown in \autoref{eq:vae_loss}.
The first term is the reconstruction loss and the second is the latent encoding regularisation loss.
The purpose of the former is to get the VAE to actually encode the data.
The latter term forces the latent space to approximate a normal distribution;
to give it a convenient shape.
More technically, the loss components are the Kullback-Leibler divergence between the approximate ($q$) and the true posterior ($p$) distributions, given datapoints ($x$) and model parameters ($\theta$), and the lower bound ($\mathcal{L}$)w on the marginal likelihood of datapoint \textit{i} respectively \cite{kingma2013auto}.

\begin{equation}
	loss_{VAE} = D_{KL}(q_{\phi}(z | x^{(i)}) \|p_{\theta}(z | x^{(i)})) + \mathcal{L}(\theta,\phi;x^{(i)})
	\label{eq:vae_loss}
\end{equation}

The configurable parameters of the VAE are the number, size and activation functions of hidden layers, the size and noise rate of the encoding layer as well as the losses and optimisers.
These parameters affect the nature of the VAE.

We are not aware of work formally defining a "good" VAE but from first principles of NNs, we can assume that the latent space must not be underfitted nor overfitted.
Avoiding underfitting means that the VAE needs to encode the features of the data.
The purpose of this is obvious:
if the network does not encode the data, its latent space has no meaning whatsoever.
Under visual inspection, an underfitted latent space appears like a spherical cloud of points around zero.
Practically, underfitting can be easily identified by whether the model's loss value descends -- i.e. the network trains.
An overfitted latent space, in turn, looks like a manifold, e.g. a line.
In this case the network learned to simply memorise the datapoints and mapped them to distinct points, rather than an area.
In the case of overfitting, the latent space encoding is likewise meaningless.



\subsection{Generative Neural Network}

The input to the GNN is the encoding of a trace produced by the VAE, with addition of small Gaussian noise $\sigma=0.1$.
Gaussian noise as input was also considered but discarded as a less interpretable; there would be no mapping from behaviours to program inputs.
The purpose of this noise is to perturb the data so that the framework explores novel behaviours.
Much like in DCGAN, the input is then upscaled with a densely connected layer(s), and reshaped to match the size of the output of the GNN (using deconvolutions with stride $>1$).
The main portion of the GNN is a stack of transposed convolutions (commonly called "deconvolutions" -- hence DCGAN) with strides one or two.
This stack of deconvolutions may also include shortcuts, i.e. residual blocks \cite{he2016deep}.
These are intended to strengthen the signal through the network thereby alleviating the vanishing gradient problem \cite{hochreiter1998vanishing}.

The output layer is of size $(str\_len_{max}, dict\_size)$.
In our experiments the maximum string length $str\_len_{max}$ was fixed at $512$ and the number of possible characters $dict\_size$ was 129:
128 ASCII characters and 0 for padding.
During training, the output layer is trained using softmax cross-entropy of one-hot encoded strings and the reproduction error of the encoded trace (\autoref{eq:gnn_loss}).
The first term $-\sum_{c=1}^{M}y_{o,c} \log(p_{o,c})$ trains the the GNN to approximate the mapping of trace encodings to strings.
It is a standard loss for training categorical classifiers \cite{Goodfellow-et-al-2016}.
The second term $|\hat{z} - z|$ aims to approximate the SUTs semantics within the GNN.
This term is minimised when the value of the encoded trace on the input $z$ to the GNN matches the \textit{true} trace $\hat{z}$ encoding that the SUT produces when executed under the generated input.
The overall loss function maps traces to strings \textit{and} tries to make those strings produce a specific behaviour.

\begin{equation}
loss_{GNN} = -\sum_{c=1}^{M}y_{o,c} \log(p_{o,c}) + |\hat{z} - z|
\label{eq:gnn_loss}	
\end{equation}

During generation, rather than simply choosing the most likely character for each index of the output string ("argmax"), the generator samples from the output probability distribution, $x = \langle x_i \sim p_i; i < str\_len_{max} \rangle$.
This sampling technique selects \textit{plausible} (rather than \textit{most likely}) characters given a learned probability distribution.

There are numerous possible parameter configurations for the GNN.
Common options include the size and the number of upsampling and deconvolutional layers, activation functions and optimisers.
In addition, fundamental architectural options had to be considered.
Namely, whether to use string reconstruction cross-entropy, trace reconstruction MSE or their combination as a loss; whether to use trace encodings or Gaussian noise as the input; and whether to sample or use argmax on the output.

\subsection{Experiments Conducted}

We conducted experiments on three SUTs that take strings with complicated grammar as input.
The first one is an XML linter called \textit{xmllint} from libxml \cite{libxml}.
The second is \textit{sparse} \cite{sparse}, a lightweight parser for C \cite{sparse}.
The last program is \textit{cJSON}, a parser for the JSON format \cite{cjson}.
Generating inputs for these programs out of thin air is a non-trivial task.
Also, since these programs validate their inputs by design, the validation process ought to be reflected in execution traces.
Execution traces that depend on syntax validation would give GNAST and AFL a good representation to work with.

For the comparative analysis against a baseline, we first ran AFL on each program for 48 hours.
This produced 3240, 11286 and 2071 queue inputs for \textit{xmllint}, \textit{sparse} and \textit{cJSON} respectively.
Of the three SUTs, AFL only found crashes in \textit{sparse}.

In GNAST's implementation, we evaluated its the performance under dozens (if not hundreds) of parameter configurations.
There was not a fixed time budget, nor a strict performance measure for each configuration.
Instead it was a trial and error exploration of the effects of various configurations with respect to the research questions.
Once the most promising structure and configuration was found, we trained GNAST on each program for 48 hours.
These trained instances of GNAST were then be used for RQs three through six.

\section{Results}\label{sec:results}

The research questions explore the nature of the proposed architecture.
The findings of our investigation show that GNAST does indeed train while maintaining diversity, that it produces sensible program inputs, and that proximity in the latent space corresponds to similarity of generated strings.
However, the result to RQ5 was negative:
we cannot generate program inputs that would trigger specific program behaviours.
Finally, GNAST does find crashes in \textit{sparse} -- the SUT in our corpus where AFL also found crashes.


\subsection{RQ1 -- Training Convergence}

GNAST trained successfully under some configurations, however it failed under others.
Unlike standard NNs, a successful training is not meant to converge arbitrarily close to zero because GNAST is ought to continually produce new data.
Instead, the convergence ought to be simply \textit{noticeable}.
Our primary definition of failure is the failure to converge, i.e. underfitting.
A secondary notion of failure is instability -- a case when the training failed with the loss reaching \verb|inf|.
In fact, these two effects correspond to a vanishing and exploding gradient problems respectively \cite{hochreiter1998vanishing, pascanu2012understanding}.

The VAE's loss failed to converge when the noise in the encoding layer was too high and when the hidden layers were very small, e.g. 8 units.
Lowering the amount of noise to $\sigma = 0.1$, using 512 or more neurons in the hidden layers along with a Leaky ReLU activation allowed the VAE to converge.

GNN's training did not converge when the structure was too weak;
there were too many layers or too few filters.
When it came to depth, more than a dozen layers did not appear to converge in a reasonable time.
Residual blocks resolved this issue however and structures of up to 64 layers converged.
With fewer than 32 filters the network also failed to converge.

Initial configurations often resulted in instability, indeed much more often than we had previously encountered in working with NNs.
We believe there are three reasons for the instability.
The first is aggressive optimisers, particularly Adam \cite{kingma2014adam}.
Adam is susceptible to instability when close to convergence \cite{wilson2017marginal}, which is a potential explanation.
The second reason is Gaussian noise in the latent layer of the VAE and on the input to the GNN.
Lowering the noise $\sigma$ to $0.1$ resolved this instance of instability.
The last cause, we suggest, is the constantly shifting dataset.
This is the most unusual reason, specific to GNAST, but not to NNs in general.
Modern optimisers like RMSProp \cite{tieleman2012lecture} and Adam both keep track of past gradients to calculate the next optimisation steps.
Adam also uses momentum.
Injecting new data throws these optimisers off:
taking a step towards what previously seemed an optimal point, and finding a completely unexpected datapoint there causes the loss to diverge.
We are not aware of work where the training dataset would be updated like in GNAST so this suggestion is speculative, and warrants further research.

The final, most promising set of hyperparameters was the following.
An RMSProp optimiser with a learning rate to $0.0001$, $42$ convolutional layers with residual connections, a kernel size of $3$, Leaky ReLu activation, batch normalisation after activation, VAE's latent encoding as input, and a categorical cross-entropy loss.

The training loss followed an unusual trajectory:
first down quickly, then up, then down slowly again.
This is not a major finding per se, but an observation of expected behaviour, given a framework that produces its own training data.
While there are few datapoints at the start of training, the model quickly learns how to reproduce them and the loss goes down.
As new datapoints are found, the loss increases, and then slowly descends again when features of the new datapoints are learnt.

The central finding of this RQ is that GNAST trains.
The positive outcome to this RQ validates the principle of GNAST and makes subsequent questions worthwhile.

\subsection{RQ2 -- Diversity During Training}\label{res:rq2}

The second property of a successful training process is that it maintains diversity.
Since GNAST produces its own training data, it is critical for the framework not to collapse into generating the same outputs over and over.
Overfitting was seen both in the VAE and the GNN.
We consider a model failing to maintain diversity if no new traces are produced over 20 epochs of generation and training.

When the VAE was too powerful, the loss tended close to zero and under visual inspection, the latent space looked like the datapoints are placed on a manifold.
This occurred when there was no noise or the hidden layers were too powerful (too many cells or layers).
In this scenario, the locality in the latent space is meaningless which rids the VAE of its purpose.
That is to say, an overfitting VAE is a useless structure with respect to ordering the execution traces -- not necessarily an immediate problem for diversity.

Making the GNN too strong would in turn result in a reduction in diversity of generated strings.
Too many filters in the deconvolutional layers, too little noise on the input or when the dropout rate was set too low -- all contributed to a diminished diversity.
We do not have a principled explanation for this, but it may be due to the GNN effectively learning to ignore the input layer and just encoding the dataset into its structure, before it manages to become sufficiently diverse and representative -- an artefact of the recursive nature of GNAST.

The fundamental architectural choice of sampling on the output, rather than taking the maximum, had a very strong effect on diversity.
In an untrained model, changes in the input propagated to the output very weakly.
That is, when feeding inputs drawn from a normal distribution to the GNN, and then taking argmax at the output, the generated strings altered very slightly, if at all.
Sampling on the output however readily produces numerous varied outputs.
This provides ample training data for GNAST while also following the learnt output distribution.

The best loss function with respect to diversity was the combination of cross-entropy on the outputs and MSE between the input to the GNN and the resulting true encoding (when the generated string was passed through the SUT and the VAE).
Using only cross-entropy, the model did produce diverse outputs initially, but then lost diversity and stabilised into preferred outputs.
Using only the encoding reproduction, the framework did not appear to converge, and new inputs were found very slowly, perhaps even coincidentally.
In combination however, the model kept producing new outputs seemingly indefinitely\footnote{We ran the model for over 96 hours and it kept generating new inputs for each SUT.}.
In addition, this combination of components for the loss function had a clear effect on crash discovery, as described below in \autoref{res:rq6}.
We cannot explain why it is the \textit{combination} of these loss components that has this property and this is a central direction for future work.


\subsection{RQ3 -- Syntactic Features}

One of the characteristics we desire of GNAST is that it ends up generating sensible program inputs.
We evaluate the "sensibleness" of the generated corpora by inspecting the most common n-grams.
This analysis is comparative because the current implementation of GNAST uses AFL's instrumentation, so it could not produce better formed strings than those made by AFL.

N-grams are sequences of characters of lengths $[3...10]$ that occur most frequently in the corpora.
Typical features found in corpora produced by AFL and GNAST are discussed below and sample snippets are presented in Listings \ref{lst:AFL_examples} and \ref{lst:GNAST_examples}.

There were similarities in n-grams across corpora generated by both AFL and GNAST.
For \textit{xmllint}, the most prevailing features are the triangular braces \verb|<| and \verb|>|.
Usually these are not properly matched however.
That is, neither tool finds that opening and closing braces ought to come in pairs.
Colon \verb|:| also appeared often.
In XML, this special character is used to denote a name prefix to resolve name conflicts.
Though \verb|:| comes up often, it is not in the correct location, i.e. within a tag name.
Another control sequence that kept appearing is the processing instruction \verb|<?| which sends the command within a tag to third party software \cite{xmlSyntaxRules}.
\textit{Sparse} is a very simple parser for C code and it only identifies basic syntactic errors.
It is therefore unsurprising that the generated strings did not appear very much like actual C code.
Two features stood out however:
braces \verb|(| and \verb|)| and the presence of semicolons \verb|;|.
Common character sequences for \textit{cJSON} included curly and square braces \verb|{|, \verb|}| and \verb|[|, \verb|]|, colons \verb|:| and line breaks \verb|\n|.
These are all special characters associated with the correct syntax of JSON.
Again though, neither AFL nor GNAST generated anything resembling well-formed JSON.
These observations were in line with the expectation of some syntactic features, but not well-formedness.

There were also differences in the corpora generated by AFL and GNAST.
The first difference is the length of generated strings.
Whilst AFL's heuristics keep the maximum length unrestricted, under the current implementation, the maximum string length of GNAST is capped at 512 characters.
Furthermore, AFL occasionally prunes the corpus by deleting slices of the strings and observing whether this changes the execution traces.
GNAST on the other hand has no preference for shorter strings, so most of the strings in its corpus are ca. 500-510 characters in length\footnote{Strings of length shorter than 512 are due to removed padding bytes in generated strings.}.
Second, strings of AFL's corpus contain very long sequences of repeated characters and character sequences.
For instance, there are input strings where the letter \verb|K| is repeated thousands of times.
Although GNAST sometimes repeats sequences as well, not nearly to such degree.
Third, strings produced by GNAST appear to be more exploratory:
where AFL finds an interesting n-gram and keeps reusing it, GNAST does not.
As a result, GNAST generated strings appear less structured.
Finally, while neither tool is restricted to only printable characters, AFL uses them more frequently than GNAST does.
To our knowledge, AFL does not have heuristics that prioritise printable characters, so we do not have an immediate explanation for this effect.

It is obvious that the generated strings are a far cry from well-formed, whether generated by AFL or GNAST.
Nonetheless, they are not random sequences of characters either;
some syntactic features are clearly identifiable.

We conclude that GNAST generates strings that appear to be comparable in nature to those made by AFL.
This means that the feedback mechanism works and that GNAST leverages it effectively to learn significant features of the SUTs' input syntax.

\begin{lstlisting}[caption={Sample snippets of input strings generated by AFL.}, label={lst:AFL_examples},captionpos=b,numbers=left,basicstyle=\small] 
// xmllint
<YY>vYY&&&aZaaaa
<b:><b:\\><
<Y:YvYYb></Y:YvY
<?xml.ver:V?VFxx
<YvHP>><?PM>
<IQQQQQQQQQQQQQQQQQQQQY><YvY<YvQQQQQY=pnQ
<\xd3\xbe>\xde[]:f\xde
((((((((()

// Sparse
[KKQKKKKKKKKmKKKKKKKK*KKKKKKKKKKKKKmK
K?;K(C){K
**/****\x00\x00\x00
//////////

// cJSON
"\n{":3 \x1c\x01}
\n{":3 \x1c\x01}\x13
+\u9'999\u+\u9A99
\end{lstlisting}

\begin{lstlisting}[caption={Sample snippets of input strings generated by GNAST.}, label={lst:GNAST_examples},captionpos=b,numbers=left,basicstyle=\small] 
// xmllint
<:p><n:/>\n
HQ%\x1f>\n#jS,u\x03\x1f<\x05UV\x05\n1
+\n&<?\n7>"<?
n=#+\n_<G:7>
[WH\x13}\x19Bc<]
<(eCQ\x05\x1bkL{GX\x01=O%Fck\x05Q)

// Sparse
i;();;gT;/*[n
if()[[*(P**nc*H.F^//)
//\x044U\x15COrm
/*Q9\x15C1\x01q
cU\x01UUv {s\x1c\x15r8\x05)!0C;\r
#if({;+;?;*;U;1>Om;

// cJSON
{*,\x04}
{\x80\x1aK:5\x0843"\x08y\x0b^\x04\x19"#\x02}
{"\x1d\x7f\x10
<@?\x086\x1f\x1d\x08\x1d\x02\x1b{=\x08=\x08}

\end{lstlisting}

\subsection{RQ4 -- Similarity of Elements in Latent space}\label{sec:res_rq4}

By design, GNAST is intended to map encodings of program executions to program inputs that triggered those executions.
The latent space encodings define a notion of similarity and this ought to be reflected in the strings generated by the GNN:
strings generated from points close to each other in the latent space should be similar, and strings from distant points dissimilar.

This property was evaluated in the following way.
Ten thousand vectors from a normal distribution are drawn, and ordered by two algorithms:
the greedy FFT algorithm from most to least representative, as well as an opposite "closest first traversal" (CFT).
CFT orders the datapoints such that the pairwise distance of each following element is minimised with respect to the existing sequence.
We then feed the first hundred elements of the FFT and CFT sequences to a trained GNN and compare the generated strings.

The expectation is that the strings from the FFT sequence ought to be less syntactically similar than those generated from CFT.
Syntactic similarity is assessed by the Jaccard index of the overlap of n-grams of lengths $[1...10]$ of each string vs. the n-grams of the whole dataset of 100 elements.
That is $\frac{|A \cap B|}{|A \cup B|}$, where $A$ is the set of n-grams in each individual string and $B$ is the set of n-grams in all 100 strings.

The mean Jaccard indices are shown in \autoref{tbl:rq3}.
Since the inputs are randomly generated, the process was repeated ten times and results averaged.
The values are not to be compared across programs, but rather between the two sequences.
They show that strings generated from adjacent input values are consistently more similar in terms of n-grams versus those generated from distant points.
This means that GNAST can generate new program inputs with a notion of relative similarity.
Such an ability is a step towards ultimately defining search strategies in a continuous Euclidean space -- one of the future prospects this work aims to enable.

\begin{table}
	\centering
	\begin{tabular}{|c|c|c|}
		\hline
		\textbf{SUT} & \textbf{FFT} & \textbf{CFT} \\
		\hline
		\textit{xmllint} & 0.199767 & 0.453304 \\
		\textit{sparse} & 0.01205251 & 0.05063868 \\
		\textit{cJSON} & 0.090320 & 0.337037 \\
		\hline
	\end{tabular} 
	\caption{Mean Jaccard similarity of the n-grams of the first 100 elements in FFT and CFT sequences. Strings generated from nearby points in the input space have a higher overlap of n-grams, as shown by the higher values of the CFT column. This means that we can control the syntactic similarity of elements generated by GNAST.}
	\label{tbl:rq3}
\end{table}

\subsection{RQ5 -- Targeting Specific Behaviours}

A further intended property of GNAST is the ability to generate program inputs that trigger a specific behaviour.
In other words, given an input $z$ to the GNN, it generates a program input $x$, which when executed by the SUT produces a trace $t$, and when $t$ is encoded, the encoding $z'$ ought to be close to $z$.

The evaluation here was straight forward.
We sampled 100 GNN inputs $Z$, passed them through the framework and looked at the cosine distances $D_{cos}(Z,Z')$.
A cosine distance of $0.0$ is perfect correlation, $2.0$ inverse correlation and $1.0$ a lack of correlation.
Results to this RQ were negative: consistently within $0.1$ of $1.0$, the actual encoding values of traces $z'$ were random with respect to the inputs $z$. So we cannot generate a string which will trigger an exact, specific behaviour.
We do not have an immediate explanation for why this is, and we intend to look into this in future work.

\subsection{RQ6 -- Finding Crashes}\label{res:rq6}

This RQ is very important for a fuzzer in the long run.
At this stage however, GNAST is a proof of concept with some limitations and numerous ideas for future work (discussed below in \autoref{sec:future_work}).
The question of whether GNAST finds crashes is therefore merely an indication of its potential -- not an evaluation of its current ability.
The only SUT where AFL discovered crashes is \textit{sparse} and GNAST found crashes in it as well.
This is not evidence of GNAST being a better fuzzer than AFL, but that the framework has the potential of being made into one, down the line.

Crash discovery was affected by the loss function of the GNN surprisingly strongly.
We observed above in \autoref{res:rq2} that the MSE component of GNN's loss is useful for maintaining diversity.
Furthermore however, without this component, GNAST did not discover crashes in \textit{sparse}.
We retrained GNAST against \textit{sparse} ten times to confirm this effect:
5 with the MSE loss, and 5 without.
In each case, it did not discover crashes without the MSE loss.
This is a somewhat baffling finding with two consequences.
First, the MSE loss is indeed important for exploration as discussed in \autoref{res:rq2}.
Second, exploration is, in this case, important for crash discovery.
At this point it is unclear why the MSE component helps exploration and thus crash discovery, and we intend to investigate this effect thoroughly in future work.

\section{Future Work}\label{sec:future_work}

The current implementation of GNAST is a proof of concept prototype.
As such, it has a number of limitations that can be investigated in future work.

First, there are just very many possible configuration options and given the novelty, there are hardly any reference systems on which to model the parameters.
A larger, more systematic ablation study should be conducted in the future.
Second, we use a generator based on transposed convolutions where the maximum string length is fixed.
Methods for allowing variable and indeed unlimited strings lengths should be investigated.
Third, the alphabet used for generating strings is the whole ASCII character set.
Introducing some prior knowledge of the target syntax, e.g. using sequences of characters or a restricted subset may improve the performance of GNAST.
Fourth, we investigated the trade-off between training convergence and diversity using fixed learning rates, learning schedules and noise levels, which might not have been optimal.
That is, a better trade-off of learning and exploration could be achieved by dynamically adjusting the learning schedules and noises.
Fifth, the expressiveness of the trace representation we lifted out of AFL is limited;
it is a somewhat crude abstraction of program behaviour.
Alternative execution traces profilers ought to be studied.
Finally, there is the negative result to RQ 5.
As said, we cannot explain this outcome at this stage.

Despite these limitations, we believe and hope that this work will be of interest to the community and inspire future research and collaboration.


\section{Conclusion}

In this paper, we present the Generative Network for Automated Software Testing (GNAST) framework.
It is intended to address multiple problems of SBST.
These include problems with the selection, design, interpretation and implementation of representations, fitness functions and input generation methods.
Furthermore, it bypasses a fundamental issue of scarcity of representative and useful training data for neural networks.

GNAST generates program inputs with a generative NN, collects the resulting execution traces, and maps them onto an n-dimensional space with an autoencoder.
The encoding imposes an ordering on executions so we may reason about similarity of executions quantitatively.
GNAST uses this encoded representation to prune redundant datapoints in order to keep the dataset maximally representative of the range of observed program behaviours.
The process is continued ad infinitum with the framework continually exploring the SUT's behaviours while learning to produce ever more useful inputs.
We believe GNAST to be the first of its kind generative neural network for automated software testing;
one based on the idea of striving towards a novel notion of diversity by leveraging a SUT as an external oracle -- much like the environment in an RL scenario.

We explored the behaviour of the proposed framework in a series of research questions.
We showed that such an architecture can be trained, and that it keeps producing new, diverse and sensible program inputs.
We also showed that GNAST's notion of execution trace similarity translates to syntactic similarity of the generated strings.
This means that we can produce new program inputs with a continuous control of their syntactic similarity.
Although we intended to, we cannot currently generate program inputs such that they would trigger an exact target behaviour.

We believe our work to be conceptually novel for SBST, NNs and their combination.
Our analysis was not a complete, systematic exploration and ablation of all the possible configurations and options.
Currently GNAST is a proof of concept which requires further work.
On the other hand, it also opens numerous directions for future research and will be hopefully met with interest in the community.

\clearpage

\bibliographystyle{ACM-Reference-Format}
\bibliography{bibliography}

\end{document}